# Quantum Optical Coherence Tomography using two photon joint spectrum detection (JS-Q-OCT)

SYLWIA M. KOLENDERSKA[1,2,*,†], FRÉDÉRIQUE VANHOLSBEECK[1,2], AND PIOTR KOLENDERSKI[3,†]

[1]*The Dodd-Walls Centre for Photonic and Quantum Technologies, New Zealand*
[2]*The Department of Physics, The University of Auckland, Auckland 1010, New Zealand*
[3]*Faculty of Physics, Astronomy and Informatics, Nicolaus Copernicus University, Grudzia̧dzka 5, 87-100 Toruń, Poland*
[*]*Corresponding author: skol745@aucklanduni.ac.nz*
[†]*These authors contributed equally to this work*

**Quantum Optical Coherence Tomography (Q-OCT) is the non-classical counterpart of Optical Coherence Tomography (OCT) - a high-resolution 3D imaging technique based on white-light interferometry. Because Q-OCT uses a source of frequency-entangled photon pairs, not only is the axial resolution not affected by dispersion mismatch in the interferometer, but is also inherently improved by a factor of square root of two. Unfortunately, practical applications of QOCT are hindered by image-scrambling artefacts and slow acquisition times. Here, we present a theoretical analysis of a novel approach that is free of these problems: Q-OCT with joint spectrum detection (JS-Q-OCT). Based on a photon pair coincidence detection as in the standard Q-OCT configuration, it also discerns, each photon pair by their wavelength. We show that all the information about the internal structures of the object is encoded in the joint spectrum and can be easily retrieved through Fourier transformation. No depth scanning is required, making our technique potentially faster than standard Q-OCT. Finally, we show that the data available in the joint spectrum enables artefact removal and discuss prospective algorithms for doing so.**

## 1. INTRODUCTION

Optical Coherence Tomography (OCT) is a high-resolution 3D imaging technique based on white-light interferometry [1]. Because it is non-destructive, non-invasive and can visualise structures at a micron scale, it found many applications in medicine [2] in the areas of ophthalmology, oncology or even dentistry, but also in industry for precision laser machining [3]. OCT provides an A-scan or depth profile of an object at one lateral point by interferometrically measuring the time of flight of light back-scattered from the structures of the object. The earliest implementation of this technique was time-domain OCT, where an A-scan is directly obtained by axially translating the reference mirror, and the signal is detected by a photodiode. The real turning point for OCT and OCT-related research was Fourier domain OCT where the reference mirror's position remains fixed and an A-scan is produced by Fourier transformation of the spectral interference. Such approach enabled achieving superior sensitivity levels [4] and much shorter acquisition times letting OCT evolve into a tool which can retrieve various types of information about the object beyond its structural characteristics [5]. For instance, OCT is able to investigate object's elasticity [6], birefringence properties [7] or can detect and characterize internal flows [8].

However, OCT still has limitations and further improvements are needed. Its depth resolution is limited by the coherence length of the light source and although it can be smaller than 1 µm for spectrally broadband sources [9], it is detrimentally affected by unbalanced chromatic dispersion in the system, which causes depth-dependent image quality degradation and contrast reduction [10]. Whereas it is fairly easy to mitigate the destructive effects of chromatic dispersion mismatch in the interferometer [11–17], the dispersion introduced by the object in ultra-high-resolution imaging is still problematic [10].

Advances in quantum interferometry led to the creation of Quantum OCT (Q-OCT) [18, 19], which is immune to dispersion imbalance and provides enhanced resolution [20]. In this method, the classical light source is replaced with a source of frequency-entangled photon pairs generated using Spontaneous Parametric Down-Conversion (SPDC) [21], and the traditional interferometer (mostly in the Michelson's configuration) – by a Hong-Ou-Mandel interferometer [22, 23]. Finally, two single photon detectors measure the rate of coincidence of photon's arrival at the two output ports of a beamsplitter. When the reference arm's length approaches the object's arm length in the interferometer, the probability of the two photons exiting from the same port of the beamsplitter increases. The coincidence rate drops and a dip appears in the signal. This dip, also known as the Hong-Ou-Mandel dip, determines the axial resolution in Q-OCT and is two times smaller than the Point Spread Function width in a classical OCT signal for the same spectral bandwidth due to the nature of frequency-entanglement of the photons propagating in the interferometer. Also, thanks to this quantum entanglement, the width of the Hong-Ou-Mandel dip is not affected by even orders of dispersion. Their cancellation is in practice enough to ensure optimal axial resolution, because, for the spectral bandwidths used in OCT, only the second order dispersion significantly contributes to the broadening of the axial Point Spread Function.

Practical applications of Q-OCT are hindered by two main issues: long acquisition times and artefacts. The former is a direct consequence of the need to axially translate the reference mirror to obtain an A-scan. This is why, Q-OCT in its current form can be viewed as the non-classical counterpart of timedomain OCT.

The major roadblock for Q-OCT is artefacts – additional dips or peaks – caused by interference of photon wavepackets reflected from the object's interfaces. Because the artefact number increases quadratically with the number of interfaces in the object, the Q-OCT signal for multilayer objects becomes cluttered with artefacts. A potential strategy to remove the artefacts was proposed in early studies on Q-OCT [18]. Having noticed that slight changes in the central frequency of the pump light makes the artefacts in the resulting A-scan, or interferogram, transition from a peak to a dip and vice versa, the authors of [18] suggested that the artefacts can be entirely removed by averaging interferograms taken for multiple pump frequencies. Seventeen years later, Graciano *et al.* [24] showed it experimentally by using a spectrally broadband light source as a pump. Because a broadband pump could be viewed as a sum of different central frequencies, the resultant interferogram is basically a coherent integration of interferograms that would be created if each of these frequencies were used separately to produce an A-scan. Such "integration" should fundamentally worsen the dispersion cancelling effects, because, as shown in Ref. [20], near-perfect dispersion cancellation is only obtained for near-perfect monochromatic pump light. However, the authors showed experimental results that suggest that the effects of dispersion, even if fundamentally not perfectly cancelled, are practically negligible. Nevertheless, their method removed the major obstacle in Q-OCT - the artefacts, and therefore, is a major milestone for Q-OCT.

Most recently the same group proposed a method based on spectrally-resolved coincidence detection [25]. It is to some extent the first implementation of a quantum equivalent of Fourier domain OCT, which offers a solution to the problem of long acquisition times in Q-OCT. Unfortunately, this method relies on manual identification of artefact peaks for the reconstruction of an A-scan, so is still far from being practical.

Classical techniques were also developed based on Q-OCT principles and show enhanced resolution and dispersion cancellation [26–39]. In the literature, they are referred to as quantum-inspired OCT and quantum-mimic or quantum-mimetic OCT and they present various levels of success in removing artefacts and achieving better sensitivity.

Here, we propose a new Q-OCT modality, where slow "time-domain" acquisition is replaced with a faster spectral detection, called Q-OCT with joint spectrum detection (JS-Q-OCT). In our method, the reference mirror position remains fixed while the joint spectrum intensity (JSI) [40] is acquired with two spectrometers (Fig. 1). Our method is based on spectrally-resolved detection of photon coincidences like in Ref. [25]. But whereas the method in Ref. [25] uses only a small portion of JSI - its diagonal - to reconstruct an A-scan, our method takes advantage of the full JSI in providing an A-scan and making it artefact-free. In Section 2, we develop a general theoretical model for Q-OCT, which we show to describe the signal in both JS-Q-OCT and after appropriate transformations - the "time-domain" equivalent. In Section 3A, we provide a general expression describing a JS-Q-OCT signal for a single-layer object. Based on this example, we discuss how different parts of JSI should be interpreted, where to look for dispersion cancellation in JSI and how the artefacts can cancel themselves out. To visualise these aspects, we provide a numerical simulation of a signal for a single-layer object. Section 3B contains a generalized expression describing a JSI for *N* layers and goes into greater detail of how the artefacts can be removed for more complicated objects. A simulation of a JSI for a double-layer object is presented to visualise the behaviour of the artefacts and the concepts behind their removal. In Section 3C, two artefact removal algorithms are proposed and then applied to the numerical data for single- and double-layer objects. Finally, the discussion and conclusions are provided in Section 4.

## 2. JOINT SPECTRUM IN QUANTUM OCT (Q-OCT)

The principle of operation of Quantum OCT with joint spectrum detection (JS-Q-OCT) is presented in Fig. 1. Just as in the case of "time-domain" Q-OCT, entangled photon pairs are created in the process of Spontaneous Parametric Down-Conversion (SPDC) in a nonlinear crystal. One photon propagates in the object placed in the object arm of the interferometer and the other one is reflected from a mirror in the reference arm. Both photons overlap at a beamsplitter and two spectrometers measure a wavelength-dependent rate of coincidence of their simultaneous detection, $C_{JSQOCT}(v_1, v_2)$ also known as joint spectrum intensity (JSI). Here $v_1$ and $v_2$ are frequencies detuned from the central frequency $\omega_0$, which is half of the central frequency of the pumping laser. For simplicity, $v_1$ and $v_2$ will be used throughout the calculations to identify the photons in a pair. The joint spectrum amplitude (JSA) $\phi(v_1, v_2)$ [41] of the photon pair:

$$\phi(v_1, v_2) = \exp\left(\frac{(v_1 - v_2)^2}{2\sigma_d^2} + \frac{(v_1 + v_2)^2}{2\sigma_p^2}\right) \quad (1)$$

leads to measurable JSI profile, $|\phi(v_1, v_2)|^2$, where $\sigma_d$ is the spectral width of the light and $\sigma_p$ depends on the SPDC crystal and spectral mode characteristics. For CW laser pumping, the width $\sigma_p$ - in the idealized case equal 0 - is very narrow and yields a joint spectrum featuring negative correlations, where the detection of photon with a lower frequency is correlated with the detection of a photon with a higher frequency. This feature guarantees the even-order dispersion cancellation for Q-OCT. In contrast to all other Q-OCT applications, we propose to use a pulsed laser pump, which generates larger spectral widths, $\sigma_p$ as the two photons' frequencies are not as well defined as in the case of a CW laser. However, a broader joint spectrum contains more information about the imaged object allowing new features in Q-OCT: purely algorithmic near-perfect dispersion cancellation and artefact removal.

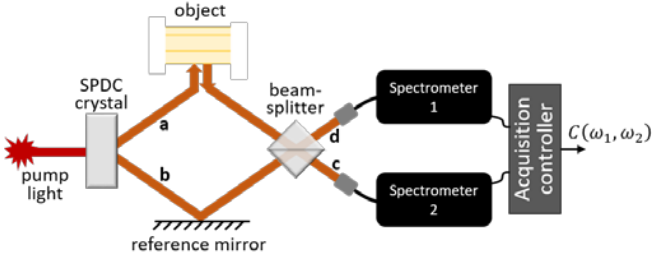

**Fig. 1.** In JS-Q-OCT, a pump generates entangled photon pairs. One photon (central frequency $\omega_1$) propagates through the object in the object arm, and the other one (central frequency $\omega_2$) is reflected from the reference mirror in the reference arm. They both overlap at a beamsplitter. Spectral relationships of the photons in a pair are measured on top of their coincidence rate in the form of joint spectrum intensity (JSI), $C_{JSQOCT}(\omega_1, \omega_2)$, by means of two spectrometers synchronized and controlled by the acquisition controller [25, 42, 43]. The paths that a photon can follow in the system are labeled with a, b, c and d.

The photons generated in the SPDC crystal propagate along paths marked as $a$, $b$, $c$ and $d$ as depicted in Fig. 1. We associate quantum-mechanical operators creating a photon with a given frequency, $\nu$, in the mode associated with a given path, $\hat{a}^+(\nu)$, $\hat{b}^+(\nu)$, $\hat{c}^+(\nu)$ and $\hat{d}^+(\nu)$, respectively. The state of the two photons just before one of them starts to propagate in the object can be written in Dirac notation as:

$$|\psi\rangle = \int d\nu_1 d\nu_2 \phi(\nu_1, \nu_2) \hat{a}^+(\nu_1) b^+(\nu_2) |0\rangle, \quad (2)$$

where $|0\rangle$ stands for a vacuum state. Next, the photon in the object arm, while propagating through the object, acquires a phase which we model with a transfer function $f(\nu)$. A photon in the reference arm is reflected from the reference mirror. To not lose generality, we assume that the reference mirror can be translated and, thus, introduces a temporal delay $\tau$. The two phase contributions modify the two photon wave function in the following way:

$$|\psi(\tau)\rangle = \int d\nu_1 d\nu_2 \phi(\nu_1, \nu_2) f(\nu_1) e^{i\nu_1\tau} \hat{a}^+(\nu_1) b^+(\nu_2) |0\rangle \quad (3)$$

Next, the two photons interact at the beamsplitter. Consequently, the final wavefunction is described in terms of creation operators associated with the output ports:

$$|\psi(\tau)\rangle = \int d\nu_1 d\nu_2 \phi(\nu_1, \nu_2) f(\nu_1) e^{i\nu_1\tau} \\ [\hat{c}^+(\nu_1) + d^+(\nu_1)][\hat{c}^+(\nu_2) - d^+(\nu_2)]|0\rangle \quad (4)$$

The measurement is based on a photon coincidence detection in the output ports $c$ and $d$. As a result, only the terms with a product of $\hat{c}^+(\nu)$ and $\hat{d}^+(\nu)$ are of interest and the wavefunction can be rewritten as:

$$|\psi(\tau)\rangle_{coinc} = \int d\nu_1 d\nu_2 f(\nu_1) e^{i\nu_1\tau} \hat{c}^+(\nu_1) d^+(\nu_2) \phi(\nu_1, \nu_2) \\ [-f(\nu_2) e^{i\nu_1\tau} + f(\nu_1) e^{i\nu_2\tau}]|0\rangle \quad (5)$$

and gives the the coincidence probability density:

$$p(\nu_1, \nu_2, \tau) = |\phi(\nu_1, \nu_2)|^2 \left| -f(\nu_2) e^{i(\nu_1)\tau} + f(\nu_1) e^{i(\nu_2)\tau} \right|^2 \quad (6)$$

This is the formula that describes both JS- and "time-domain" Q-OCT. In the "time-domain" Q-OCT case, the detection system does not allow for spectrally resolved measurements, so the photon count statistics is measured instead:

$$\begin{aligned} C_{QOCT}(\tau) &= \int p(\nu_1, \nu_2, \tau) d\nu_1 d\nu_2 \\ &= \int |\phi(\nu_1, \nu_2)|^2 (|f(\nu_1)|^2 + |f(\nu_2)|^2) d\nu_1 d\nu_2 \\ &\quad - \int |\phi(\nu_1, \nu_2)|^2 2\Re\{f(\nu_1) f^*(\nu_2) e^{i(\nu_2-\nu_1)\tau}\} d\nu_1 d\nu_2 \\ &= \Lambda_0 + \Lambda(\tau), \end{aligned} \quad (7)$$

where $\Lambda_0$ is a self-interference contribution to the measured interferogram, and $\Lambda(\tau)$ is a cross-interference term, which contains information about the object's structure. The equation above reproduces the result of Ref. [18].

In the case of JS-Q-OCT the time delay, $\tau$, is fixed as the reference mirror does not move. The measured signal is a two-dimensional modulated joint spectrum:

$$C_{JSQOCT}(\nu_1, \nu_2) = p(\nu_1, \nu_2, \tau) = |\phi(\nu_1, \nu_2)|^2 M(\nu_1, \nu_2) \quad (8)$$

where $M(\nu_1, \nu_2) = |f(\nu_1)|^2 + |f(\nu_2)|^2 - 2\Re\{f(\nu_1) f^*(\nu_2)\}$ is the modulation component and incorporates dispersion-free structural information about the object.

## 3. JS-Q-OCT SIGNAL

To better understand how even-order dispersion cancellation is achieved and where the artefacts in a JS-Q-OCT signal stem from, we provide, in Subsection A, a detailed expression describing a JS-Q-OCT signal for a single-layer object with given parameters together with a simulation of a signal for a 50-$\mu$m thick quartz, and, in Subsection B, a general expression of the signal for $N$ layers with given parameters together with a simulation of a double-layer object: a 50-$\mu$m thick quartz on top of a 120-$\mu$m thick BK7. In Subsection C, we propose two algorithms to retrieve artefact-free depth profiles for these two types objects.

### A. Single-layer object

Let's first assume that the object consists of one dispersive layer of thickness $z_2$ placed at a distance $z_1$ from zero optical path difference (OPD) point. The total phase acquired in such an object will be expressed as:

$$f(\nu) = R_1 e^{iz_1(\beta_0 + \beta_1 \nu)} + R_2 e^{iz_1(\beta_0 + \beta_1 \nu)} e^{iz_2(\beta_0 + \beta_2 \nu + \beta_{22} \nu^2)}, \quad (9)$$

where $\beta_0$ is the wavenumber of light in air and $\beta_1$ is the inverse of the group velocity of light propagating in air. $\beta_2$ is the inverse of the group velocity of light propagating in the dispersive layer and $\beta_{22}$ is the Group Velocity Dispersion (GVD) coefficient for this layer. The reflectivity of the layer's interfaces are $R_1$ and $R_2$. $f(\nu)$ will basically be a transfer function of an object that consists of a $z_1$-thick layer of air and a $z_2$-thick dispersive layer beneath it.

Substituting Eq. 9 in Eq. 6 gives the most generic form of the probability function for a one-layer object. Making all the necessary transformations, the modulation component $M$ becomes a sum of cosines:

$$\frac{1}{2}M(\nu_1, \nu_2) = -R_1^2 \cos\left[2z_1\beta_1(\nu_1 - \nu_2)\right]$$
$$- R_2^2 \cos\left[2z_1\beta_1(\nu_1 - \nu_2) + 2z_2(\beta_2 + \beta_{22}(\nu_1 + \nu_2))(\nu_1 - \nu_2)\right]$$
$$+ R_1R_2 \cos\left[2z_2(\beta_0 + \beta_2\nu_1 + \beta_{22}\nu_1^2)\right]$$
$$+ R_1R_2 \cos\left[2z_2(\beta_0 + \beta_2\nu_2 + \beta_{22}\nu_2^2)\right]$$
$$- R_1R_2 \cos\left[2z_1\beta_1(\nu_1 - \nu_2) + 2z_2(\beta_0 + \beta_2\nu_1 + \beta_{22}\nu_1^2)\right]$$
$$- R_1R_2 \cos\left[2z_1\beta_1(\nu_1 - \nu_2) + 2z_2(\beta_0 + \beta_2\nu_2 + \beta_{22}\nu_2^2)\right]$$
$$+ (R_1^2 + R_2^2). \quad (10)$$

The first two terms convey information about the structure of the object. The next two contribute a "stationary" artefact peak that is placed at a fixed distance from 0 OPD, regardless of the OPD between the object and the reference arm mirror since the modulation is independent of $z_1$. The last two terms correspond to an "instationary" artefact peak that is always positioned midway between two interfaces and also appears in the time-domain Q-OCT. An example of such JSI modulation is shown in Fig. 2(a).

A perfect even-order dispersion cancellation occurs for photons with perfect negative correlations, so when $\nu_2 = -\nu_1 = \nu$. Eq. 10 is reduced to a one-dimensional function with four terms which Fourier transform to four peaks:

$$\frac{1}{2}M(-\nu, \nu) = -R_1^2 \cos\left[4z_1\beta_1\nu\right]$$
$$- R_2^2 \cos\left[4z_1\beta_1\nu + 4z_2\beta_2\nu\right]$$
$$+ 2R_1R_2 \cos\left[2z_2(\beta_0 + \beta_2\nu + \beta_{22}\nu^2)\right]$$
$$- 2R_1R_2 \cos\left[4z_1\beta_1\nu + 2z_2(\beta_0 + \beta_2\nu + \beta_{22}\nu^2)\right]$$
$$+ (R_1^2 + R_2^2). \quad (11)$$

It can be seen here that the first two terms corresponding to the structural information of the object are now dispersion free as there the GVD coefficient, $\beta_{22}$, vanished. In practice, negative correlations correspond to the main diagonal of the JSI, which is illustrated in Fig. 2(a) with red line A.1. It means that a near perfectly dispersion-cancelled signal can be easily retrieved with a simple post-processing procedure of taking the main diagonal of the acquired JSI and Fourier transforming it.

Off-diagonal spectra in the JSI correspond to the condition $\nu_2 = -\nu_1 + \Delta\nu = \nu$, where $\Delta\nu$ is a shift from the main diagonal, and can be used to cancel out the terms in the spectrum that lead to artefact peaks after Fourier transformation. By analysing the second pair of terms in Eq. 10, it can be seen that by varying $\nu_1$ through the change of the spectral shift $\Delta\nu$ while keeping $\nu_2$ constant, the amplitude of the resultant cosine term will change and, consequently, will the height of the artefact peak that is formed by Fourier transformation. The change of height is continuous and oscillates between 0 and $2R_1R_2$. As a result, there are values of $\Delta\nu$, for which the two terms cancel each other out making the artefact peak vanish entirely. The same analysis applies to the third pair of cosine terms in Eq. 10. The second artefact, which is a result of the existence of the third pair of cosines, will disappear for the same values of $\Delta\nu$. The off-diagonals also represent near-perfect dispersion-free signals, but for different central frequencies. If we consider the frequencies,

$$\omega_1 = \omega_0 + \nu_1,$$
$$\omega_2 = \omega_0 + \nu_2, \quad (12)$$

rather than the frequency detunings, $\nu_1$ and $\nu_2$, while describing the photons in a pair, the diagonal spectrum representing negative correlations ($\nu_2 = -\nu_1 = \nu$) will correspond to photons with frequencies

$$\omega_1 = \omega_0 + \nu,$$
$$\omega_2 = \omega_0 - \nu. \quad (13)$$

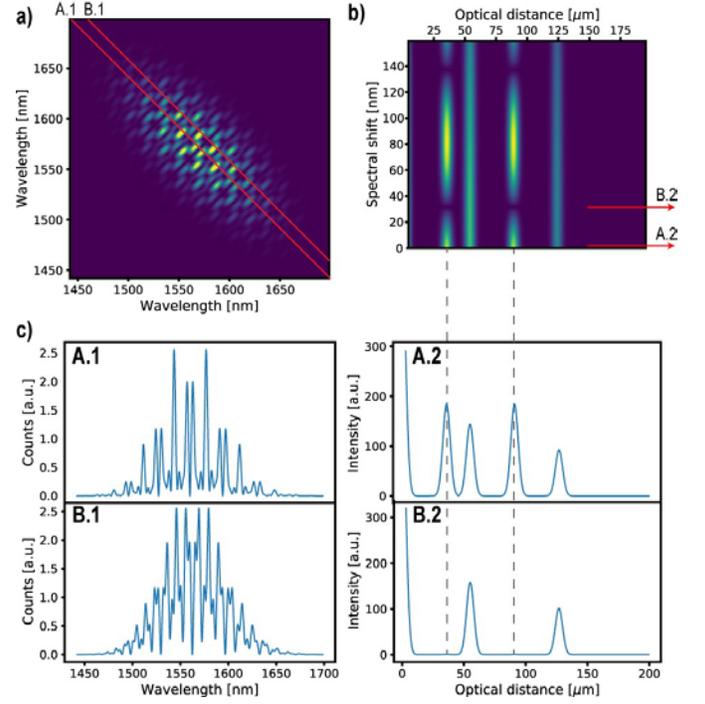

Fig. 2. (a) JSI corresponding to a 50 μm thick quartz positioned 55 μm away from 0 OPD. (b) Fourier transforms of 200 consecutive off-diagonals of the JSI show that the height of the artefact peaks goes to 0 for certain spectral shifts. (c) The main-diagonal spectrum - A.1 - is Fourier transformed to obtain an A-scan - A.2 - with two artefact peaks. An off-diagonal spectrum can be found at 31.2 nm spectral shift - B.1 - which when Fourier transformed gives an artefact-free A-scan - B.2.

For the off-diagonal spectra, the condition can be rewritten to $\nu_2 - \frac{\Delta\nu}{2} = -\nu_1 + \Delta\nu/2 = \nu$. Then, the frequencies of photons can be rewritten as:

$$\omega_1 = \omega_0 + (\nu + \Delta\nu/2) = (\omega_0 + \Delta\nu/2) + \nu$$
$$\omega_2 = \omega_0 + (-\nu + \Delta\nu/2) = (\omega_0 + \Delta\nu/2) - \nu. \quad (14)$$

Therefore, an off-diagonal spectrum corresponding to the spectral shift $\Delta\nu$ in JSI represents a spectrum for negatively correlated photons around the central frequency $\omega_0 + \Delta\nu/2$, which is illustrated in Fig. 2(a) with the red line B.1.

Fourier transformation of such a diagonal will give an A-scan with altered optical distances, because the coefficients $\beta_1$ and $\beta_2$ all depend on the central frequency. This explains why in time-domain Q-OCT, the narrower the pump spectrum, the better the dispersion cancellation [20]. For a broadband pump, the JSI expands in the anti-diagonal direction and as a result, contains a range of off-diagonals that produce depth profiles exhibiting varying optical thicknesses for a given layer. The depth profile measured in "time-domain" corresponds to a coherent sum of such off-diagonals and therefore will show a dispersively broadened layer together with a slight decrease in the resolution. On the other hand, due to this coherent averaging over different off-diagonals, the artefacts are automatically cancelled out as shown by Graciano et al. [24].

Figure 2 presents numerical data obtained for a single layer of 50-μm thick quartz positioned 55 μm away from 0 OPD with a group refractive index of 1.46 and a GVD coefficient, $\beta_2 = -28$ fs$^2$/mm. In the simulation, we assumed that the photon pair source generates JSI with $\sigma_p = 32.9$ THz and $\sigma_d = 15$ THz which results in an FWHM of 100 nm in the diagonal direction and 45.8 nm in the anti-diagonal direction. The photons are centred at 1560 nm. As expected, the Fourier transform of the main diagonal of the JSI (Fig. 2c, graph A1) incorporates two additional, artefact peaks (Fig. 2c, graph A2). To better visualise the behaviour of the artefact peaks, we Fourier transformed 200 consecutive off-diagonals from the right-hand side of the JSI, and displayed them one on top of another in Fig. 2b. The spectral distance between two adjacent off-diagonals is 0.8 nm in the horizontal direction. The intensity of the artefact peaks oscillates as the spectral shift increases: it drops down to 0 after 31.2 nm, then reaches a maximum at 81.7 nm and drops down to 0 again at 137.8 nm. The off-diagonal corresponding to 31.2 nm shift, presented in Fig. 2c on graph B1 and its Fourier transform, shows an artefact-free A-scan on graph B2.

**B. Multi-layer object**

For an object consisting of $N$ dispersive layers, the transfer function, $f(\nu)$, has extra terms corresponding to the additional layers and can be expressed as a sum of products of reflectivity coefficients, $R_n$, and phase factors:

$$f(\nu) = \sum_{n=1}^{N} R_n \exp\left(i \sum_{m=1}^{n} z_m(\beta_0 + \beta_m \nu + \beta_{2m}\nu^2)\right), \quad (15)$$

where $z_m$ is the geometrical thickness of the m-th layer, $\beta_m$ is the inverse of its group velocity, and $\beta_{2m}$ stands for its GVD coefficient.

Therefore, the modulation term, $M(\nu_1, \nu_2)$, can be generalised:

$$\begin{aligned}
\frac{1}{2}M(\nu_1,\nu_2) &= \sum_{n=1}^{N} R_n^2 \\
&- \sum_{n=1}^{N} R_n^2 \cos\left(\sum_{k=1}^{n} 2z_k(\beta_k + \beta_{2k}(\nu_1+\nu_2))(\nu_1-\nu_2)\right) \\
&+ 2\sum_{n=1}^{N-1}\sum_{m=n+1}^{N} R_n^2 R_m^2 \cos\left(\sum_{k=n+1}^{m} 2z_k(\beta_0 + \beta_k\nu_1 + \beta_{2k}\nu_1^2)\right) \\
&+ 2\sum_{n=1}^{N-1}\sum_{m=n+1}^{N} R_n^2 R_m^2 \cos\left(\sum_{k=n+1}^{m} 2z_k(\beta_0 + \beta_k\nu_2 + \beta_{2k}\nu_2^2)\right) \\
&- 2\sum_{n=1}^{N-1}\sum_{m=n+1}^{N} R_n^2 R_m^2 \cos\Big(\sum_{k=1}^{n} 2z_k(\beta_k + \beta_{2k}(\nu_1+\nu_2))(\nu_1-\nu_2) \\
&\qquad + \sum_{k=n+1}^{m} 2z_k(\beta_0 + \beta_k\nu_1 + \beta_{2k}\nu_1^2)\Big) \\
&- 2\sum_{n=1}^{N-1}\sum_{m=n+1}^{N} R_n^2 R_m^2 \cos\Big(\sum_{k=1}^{n} 2z_k(\beta_k + \beta_{2k}(\nu_1+\nu_2))(\nu_1-\nu_2) \\
&\qquad + \sum_{k=n+1}^{m} 2z_k(\beta_0 + \beta_k\nu_2 + \beta_{2k}\nu_2^2)\Big].
\end{aligned} \quad (16)$$

For negative correlations, $\nu_1 = -\nu_2 = \nu$, the modulation term is simplified further:

$$\begin{aligned}
\frac{1}{2}M(-\nu,\nu) &= \sum_{n=1}^{N} R_n^2 - \sum_{n=1}^{N} R_n^2 \cos\left(\sum_{k=1}^{n} 4z_k\beta_k\nu\right) \\
&+ 2\sum_{n=1}^{N-1}\sum_{m=n+1}^{N} R_n^2 R_m^2 \cos\left(\sum_{k=n+1}^{m} 2z_k(\beta_0 + \beta_k\nu + \beta_{2k}\nu^2)\right) \\
&- 2\sum_{n=1}^{N-1}\sum_{m=n+1}^{N} R_n^2 R_m^2 \cos\Big(\sum_{k=1}^{n} 4z_k\beta_k\nu \\
&\qquad + \sum_{k=n+1}^{m} 2z_k(\beta_0 + \beta_k\nu + \beta_{2k}\nu^2)\Big).
\end{aligned} \quad (17)$$

The second term in Eq. 17 corresponds to the dispersion-free structure of the object, the third term - to the "stationary" artefacts that clutter the area around 0 OPD, and the last one - to the artefacts that appear midway between every two interfaces. Using Eqs. 11 and 17, it can be observed that, for every layer, so for every two interfaces in the object, there are two artefacts present in the Fourier transform.

To visualise this problem, we generated numerical data for a double-layer object positioned 150 μm away from 0 OPD and consisting of a 50 μm thick quartz on top of a 120 μm thick BK7. The group refractive index of BK7 is 1.52 and its GVD is -24.6 fs$^2$/mm. The parameters of the simulation were the same as for the single-layer object. The Fourier transformation of 200 consecutive off-diagonals of the JSI were calculated (Fig. 3a), and displayed one on top of another (Fig. 3b). As expected, there are three artefacts positioned in the vicinity of the 0 OPD point and three artefacts positioned in the object itself. The intensity of each pair of these artefacts evolves differently with the spectral shift as depicted in Fig 3c. Graphs A.1 and A.2 present the main-diagonal spectrum and its Fourier transform, which is so cluttered with artefact peaks that it is not possible to discern the original three interfaces (marked as 1, 2 and 3 in Fig. 3c for clarity). The pair of artefacts corresponding to interfaces 1 and 3 disappear when the off-diagonal at a spectral shift of 18.4 nm is taken from the JSI (graph B.1) and Fourier transformed (graph B.2). In a similar way, off-diagonals at spectral shifts of 31.2 nm and 34.4 nm (graph C.1 and D.1) produce A-scans where the artefacts for interfaces 1 and 2, and 2 and 3 are suppressed (graphs C.2 and D.2).

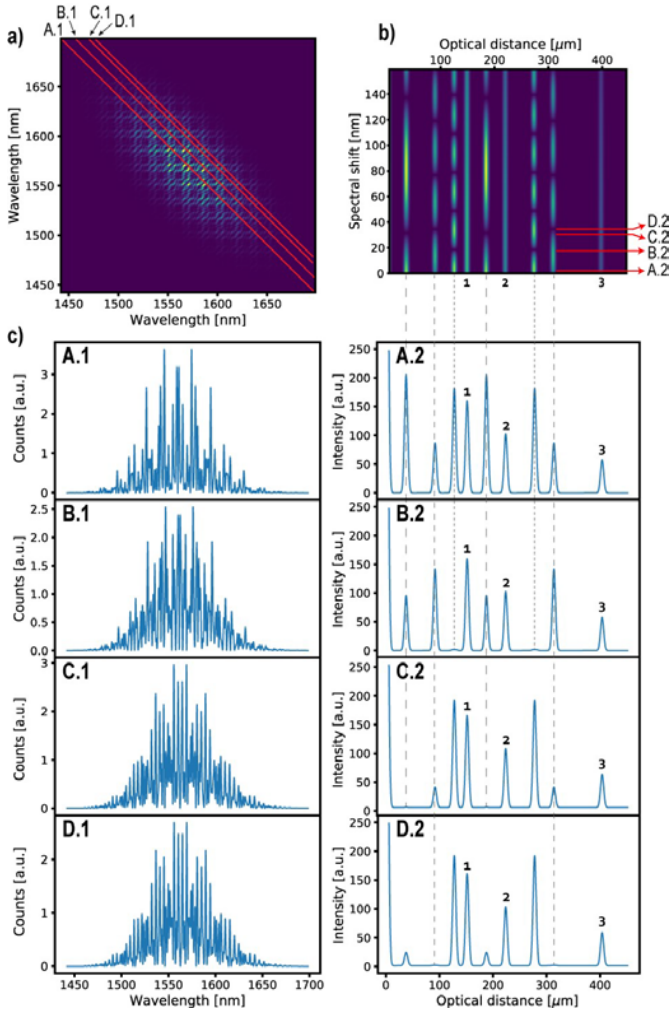

**Fig. 3. (a)** JSI corresponding to a double-layer object: a 50-μm thick quartz on top of a 120-μm thick BK7 positioned 150 μm away from 0 OPD. **(b)** Fourier transforms of 200 consecutive off-diagonals of the JSI show that the intensity of each pair of artefact peaks drops to 0 for different spectral shifts. **(c)** The main-diagonal spectrum - A.1 - is Fourier transformed to obtain an A-scan - A.2 - which is completely scrambled by artefacts. There is one pair of artefacts per every two interfaces. Their intensity drops to zero for different spectral shifts: 18.4 nm for interfaces 1 and 3 as depicted on graphs B.1 and B.2, 31.2 nm for interfaces 1 and 2 as on graphs C.1 and C.2, and 34.4 nm for interfaces 2 and 3 as on graphs D.1 and D.2.

**C. Artefact suppression**

An artefact-free A-scan is obtained by calculating the minimum values for every column in an FFT stack, which is created when consecutive off-diagonal spectra are taken from JSI and Fourier transformed, see Fig. 2 and Fig. 3. This simple algorithm was applied to the numerical data sets for single- and double-layer objects and allowed for the removal of the artefacts from the original A-scans, see Fig. 4. This algorithm does not suppress the artefacts completely. Small traces are left where the original artefacts were. These imperfections are a consequence of the fact that an FFT stack is a discrete set of Fourier transforms for which the height of the artefact peaks may not always reach a perfect zero.

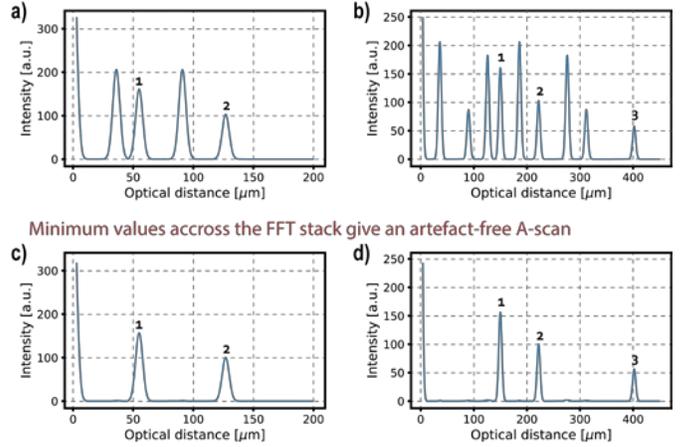

**Fig. 4. (a)** and **(b)** A-scans obtained by Fourier transformation of the main-diagonal of JSI for a single- (Fig. 2c, graph A.2) and double-layer objects (Fig. 3c, graph A.2), respectively. A procedure of taking minimum values from every column of an FFT stack presented in Fig. 2b and Fig. 3b returns artefact-free A-scans in **(c)** and **(d)**. Numbers 1 and 2 in **(a)** and 1, 2 and 3 in **(b)** mark the peaks corresponding to the structure of the object

However, this simple algorithm is not universal and fails to recover the real structure of the object in two situations. In the first one, the object's structure overlaps with the stationary artefacts and coincide with one or more instationary artefact peaks. Such a case is presented in Fig. 5a, where an object consisting of a 50-μm thick layer of quartz and 50-μm thick layer of BK7 is placed 75 μm away from 0 OPD. In this figure, the overlap of structure and artefact peaks is circled in brown. The procedure of extracting minimum values from the FFT stack results in the suppression of the structure peak as depicted in Fig. 5c. This figure also shows that a structure peak can be suppressed if it overlaps with an instationary artefact. This situation can occur when there are two layers in the object with similar optical thicknesses just as it is in the presented example where the optical thickness of quartz equals 50 μm x 1.46 = 73 μm and the optical thickness of BK7 - 50 μm x 1.52 = 76 μm.

The problem of the overlap with stationary artefacts can be easily solved by increasing the OPD between the arms of the interferometer, so by shifting the object in the A-scan, see Fig. 5b, where the object was moved from an OPD of 75 μm to 140 μm. Of course, taking the minimum values from the FFT stack will still remove the structure peak which takes the same place as one of the instationary artefact peaks (Fig. 5d). In such a case, an improved algorithm is used. This algorithm first searches the FFT stack for $N$ transforms for which one of the N stationary artefact peaks' intensity drops to 0. An artefact-free A-scan is then obtained by multiplying the $N$ found transforms and taking the $N$-th root of it (both operations done element-wise). Fig. 5e presents an A-scan, where all the artefacts where suppressed without simultaneous suppression of the structure of the object.

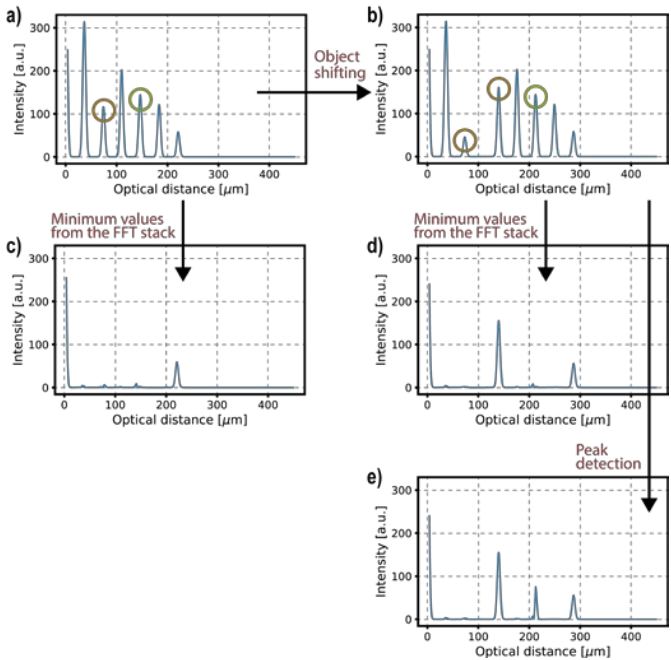

**Fig. 5. (a)** Structure peaks overlap with the stationary (brown circle) and instationary artefacts (dark green circle) resulting in six, instead of nine, peaks in the A-scan. Two stationary peaks overlap, because both layers in the object have the same optical thickness. **(b)** When the object is moved away from the stationary artefacts, an artefact peak is separated from a structure peak (brown circles). One instationary artefact still overlaps with a structure peak (dark green circle). **(c)** and **(d)** A simple procedure of taking minimum values from the FFT stacks suppresses artefacts as well as some of the object's structure **(e)** An improved algorithm keeps the object's structure almost intact while suppressing the artefacts.

Although this algorithm is also a basic one and therefore does not guarantee a perfect artefact removal, it suppresses them well enough leaving only a depth profile of the object.

## 4. SUMMARY

We have presented a study of JS-Q-OCT, a new implementation of Quantum OCT with joint spectrum detection. In JSQ-OCT, the photon pair coincidence measurement is combined with its spectral discrimination providing a two-dimensional joint spectrum intensity signal. To get an A-scan, the main diagonal of the joint spectrum intensity is taken and Fourier transformed. Both the theoretical analysis and the simulations show that the resultant A-scan contains artefacts - additional peaks that do not correspond to the structure of the object - and that these artefacts are more numerous than in the previous implementation of Q-OCT; for a raw A-scan in JS-Q-OCT, there are two artefacts per every two structural interfaces in the object. This result agrees with the experimental results for a single- and double-layer objects presented in [25]. For a single-layer object there are 4 peaks instead of 2, and for a double layer object 9 peaks instead of 3, which generalises into $n^2$ for $n$ interfaces assuming that a structural peak does not overlap an artefact or no layers have similar thicknesses.

Further analysis shows that each artefact in the pair behaves differently. One - which is positioned midway between two interfaces - is a direct equivalent of the artefact that is also present for Q-OCT signals. The second one is a result of the spectral detection and is always placed at a fixed position which is the optical (dispersive) thickness of the layer the artefact corresponds to. This behaviour is also observed in the experimental results of [25].

In JS-Q-OCT, the artefacts can be removed if the remaining part of the joint spectrum is used. The Fourier transforms of the off-diagonals of the joint spectrum provide A-scans where the height of the artefact peaks changes and at a certain points drops to zero. It should be pointed out here that an artefact peak can overlap with a structural peak, which consequently leads to the drop to zero for both. In this paper, we proposed two simple algorithms for artefact removal. The first, and the most basic one, assumes that there is no overlap between the artefact peaks and the structural peaks, whereas the second one does not makes this assumption and provides solution for more complicated experimental scenarios.

JS-Q-OCT is a true spectral equivalent of the existing "time-domain" Q-OCT and is able to produce artefact-free A-scans with a near-perfect dispersion cancellation. It shows a lot of promise in terms of acquisition time and mechanical stability, because it does not require any moving elements in the setup. Future work for this method should mainly focus on the development of more efficient artefact removal algorithms and fast and efficient spectral quantum detection schemes.

## FUNDING INFORMATION

This research was supported by Frédérique Vanholsbeeck's Marsden fund contract number UoA1509 and Sylwia Kolenderska's New Ideas fund provided by Dodd-Walls Centre for Photonic and Quantum Technologies. PK acknowledges financial support by the Foundation for Polish Science (FNP) (project First Team co-financed by the European Union under the European Regional Development Fund).

## DISCLOSURES

**Disclosures.** The authors declare no conflicts of interest.